\documentclass[a4paper]{article}
\usepackage{amsmath}
\usepackage{graphicx}

\begin{document}

\title{The Influence of Social Media Writing on Online Search Behavior for Seasonal Events: The Sociophysics Approach}

\author{\textbf{Nozomi Okano}, \textbf{Masaru Higashi} and \textbf{Akira Ishii} \\ 
 Faculty of Engineering, Tottori University, Koayam, Tottori 680-8552, Japan\\
}

\maketitle


\begin{abstract}
    Using seasonal topics as the study subject, in this study, we focus on the timing gap between social media writing and online search behavior. To conduct our analysis, we used the mathematical model of search behavior, comprising the sociophysics approach. The seasonal topics selected were St.Valentine's Day, Halloween and New Year countdown. We also picked up the event like Christmas and Halloween. We analyzed the influence of blogs and Twitter on search behavior and found a deviation of interest in terms of timing. We also analyzed Japanese seasonal event of eating Eho-maki in February 3 and eels at the day of the ox in midsummer.
\end{abstract}%

\section{Introduction}

Some people believe that Twitter and Blogs are enthusiastically posted, but most people are supposed to search on the Internet. Therefore, analysis of the search behavior of people of society is very important in grasping social movements. In this research, we analyze theoretically by the idea of social physics, how television information, net information, or Blog, Twitter etc. influences people's search behavior.

The subjects of the analysis are seasonally limited dates, and many people participate in the event. Specifically, Valentine's Day, Halloween, Christmas, and New Year's Countdown. In addition, in the habit peculiar to Japan, we also analyzed Eho-maki, which eats scroll sushi on February 3, and an event to eat eel on the Midsummer Occasion Day of Summer (the day of ox in midsummer). In particular, Specifically, Valentine's Day, Halloween, Christmas, and New Year's Countdown last for only one day and thus, it is easier to analyze the rising excitement and subsequent wane in interest. In many European countries, people return to their homelands for Christmas and enjoy a long Christmas break. In constast, in Japan, however, Christmas is a one-day event in which young people have Christmas parties with friends but do not return home. In Japan, therefore, it is not until the New Year’s vacation of the following week that people return to their hometowns. Thus, in Japan, Christmas is very similar to Halloween.

\begin{figure}[h]
\begin{center}
\includegraphics[width=6cm]{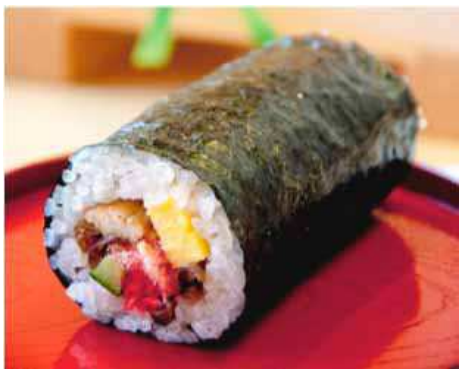}
\caption{Eho-maki in Japan.}
\label{eho-maki}
\end{center}
\end{figure}

\begin{figure}[h]
\begin{center}
\includegraphics[width=6cm]{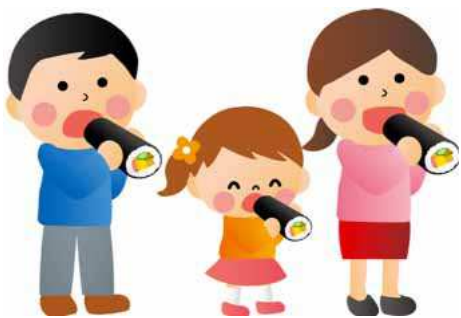}
\caption{How to eat eho-maki in Japan.}
\label{eho-maki2}
\end{center}
\end{figure}

In Japan, Feb.3 is the traditional day to eat “Eho-maki”. Eho-maki are thick sushi rolls 
shown in fig.\ref{eho-maki} which is believed to bring good fortune if eaten while facing the year’s “Eho” (good luck direction of the year) like fig.\ref{eho-maki2}. Not all, but many Japanese people have Eho-maki at this day.

\begin{figure}[h]
\begin{center}
\includegraphics[width=6cm]{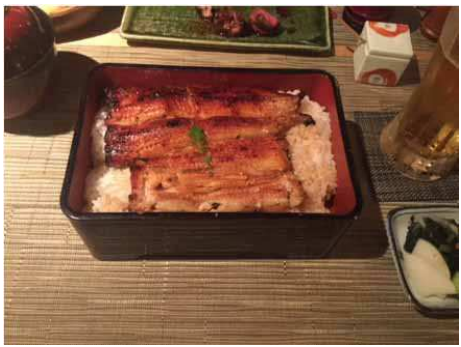}
\caption{Typical cooking of eel in Japan.}
\label{unagi}
\end{center}
\end{figure}

On the day of the ox in midsummer Japanese have a custom to eat eel which started in the Edo period, 18th century. Eel is a popular food for Japanese people, and it is expensive, so eating eel on the day of the ox in midsummer once a year is a big concern. On  the day of the ox in midsummer  Japanese have a custom to eat eel which started in the Edo period. The day of the ox, which is named after one of the twelve animals of the Chinese zodiac. According to one legend, long before the scientific reasons were established, in the 1700s well-known scholar Gennai Hiraga came up with the custom as part of a marketing ploy to boost limp summer sales when the owner of a struggling eel store asked the wise man for some business advice. In fig.\ref{unagi}, we show the typical cooking of eel in Japan for  the day of the ox in midsummer.

Thus, this study examines the peaks and falls in interest in these time-limited events, focusing on the medium used to perform searches and on what cohort of the population perform those searches. The mathematical model of search behavior is used for the analysis\cite{Ishii2018,Okano} .

\section{Theory}

In the theory of search behavior \cite{Ishii2018,Okano} , the interest and concern on a certain topic can be calculated using a mathematical model of differential equations. Here, we introduce I(t) as the interest or concern on a certain topic. We construct a mathematical model based on the mathematical model for the hit phenomenon within a society presented as a stochastic process of interactions of human dynamics in the sense of many body theory in physics \cite{Ishii2012a,Ishii2017}. As in the model in \cite{Ishii2012a,Ishii2017}, we assume that the intention of humans in a society is affected by the three factors: advertisement, communication with friends, and rumors. Advertisements act as external forces; communications with friends are a form of direct communication and its effect is considered as interaction with the intention of friends. The rumor effect is considered as the interaction among three persons and a form of indirect communication as described. In the model, we use only the time distribution of advertisement budget as an input, and word-of-mouth (WOM) represented by posts on social network systems is the observed data for comparison with the calculated results. The parameters in the model are adjusted by the comparison with the calculated and observed social media posting data. 

According to \cite{Ishii2012a,Ishii2017}, we write down the equation of purchase intention at the individual level  $I_i(t)$ as

 \begin{equation}
\frac{dI_i(t)}{dt} = \sum_{\xi} c_{\xi}A_{\xi}(t) - a I_i(t) + \sum_j d_{ij} I_j(t) + \sum_j \sum_k p_{ijk} I_j(t) I_k(t)
 \end{equation}

where t is the time, $d_{ij}$, $p_{ijk}$, and $f_i(t)$ are the coefficient of the direct communication, the coefficient of the indirect communication, and the random effect for person i, respectively\cite{Ishii2012a}.  The advertisement and publicity effects are include in $A_{\xi}(t)$ which is treated as an external force. The index $\xi$ means sum up of the multi media exposures.  
Word-of-mouth (WOM) represented by posts on social network systems like blog or twitter is used as observed data which can be compared with the calculated results of the model. The unit of time is a day. 

Here, it is assumed that the height of interest $I(t)$ of people attenuates exponentially. Although it is known that this is known to occur in movies and the like \cite{Ishii2012a}, attention such as events and anniversaries is known to attenuate by a power function. \cite{Sano2013a, Sano2013b} In the case of social interest, we attenuate the intermediate between the exponential function and the power function \cite{Ishii-Koyabu}, but here we simply adopt exponential decay.

We consider the above equation for every consumers in the society. Taking the effect of direct communication, indirect communication, and the decline of audience into account, we obtain the above equation for the mathematical model for the hit phenomenon. Using the mean field approximation, we obtain the following equation as equation for averaged intention in the society. The derivation of the equation is explained in detail  in ref.\cite{Ishii2012a}. 

\begin{equation}
\label{eq:eq13}
\frac{dI(t)}{dt} = \sum_{\xi} c_{\xi}A_{\xi}(t) + (D-a) I(t) + P I^2(t)
 \end{equation}

This equation is the macroscopic equation for the intention of whole society. Using this equation, our calculations for the Japanese motion picture market have agreed very well with the actual residue distribution in time \cite{Ishii2012a}. The advertisement and publicity effects are obtained from the dataset of M Data and the WOM represented by posts on social network systems are observed using the system of Hottolink.  We found that the indirect communication effect is very significant for huge hit movies. 

\subsection{Extension to include Twitter and Blog} 

In the new mathematical model for search behavior, we use daily blog and Twitter postings as the external force. Therefore, we extend the above mathematical model for hit phenomena to include the effects of Twitter and blog as external field as follows.

\begin{eqnarray}
\label{eq:eq14}
\frac{dI(t)}{dt} &=& C_{TV}A_{TV}(t) \nonumber \\
&+& C_{NetNews}A_{NetNews}(t) + C_{Twitter}A_{Twitter}(t) \nonumber \\
&+& C_{blog}A_{blog}(t) + (D-a) I(t) + P I^2(t)
\end{eqnarray}

In the above equation(\ref{eq:eq14}),  $I(t)$ is the intention to search a certain topic using Google Trend and $C_{TV}$, $C_{NetNews}$, $C_{Twitter}$ and $C_{blog}$ correspond to the strength to influence willingness to search the certain topic. $(D-a) I(t) + P I^2(t)$ correspond to the direct and indirect communications in the previous mathematical model for hit phenomena. In the model of the present paper, these terms correspond to the direct and indirect effect of searching the topic by other people. 

On the real calculation, we use  advertisement time data on TV from M Data co. ltd. and the internet news site data, daily Twitter posting data and daily blog posting data on the certain topic from Hottolink co.ltd. The daily search data comes from Google Trend as the reference of our calculation. 

The parameters $C_{TV}$, $C_{NetNews}$, $C_{Twitter}$, $C_{blog}$, $D$ and $P$ are determined in similar way as the previous model by using the metropolis-like Mote Carlo method as noted in the previous paper\cite{Ishii2012a, Ishii2017}. We define here "R-factor" to check the correctness of the adjustment of parameters.

\begin{equation}
R = \frac{\sum_i (f(i)-g(i))^2}{\sum_i (f(i)^2 + g(i)^2)},
\end{equation}

where  $f(i)$ and $g(i)$ correspond to the calculated $I(t)$ and the observed number of Google Trend data. The R-factor is originally defined by J B Pendry\cite{Pendry} to adjust the positions of atoms on surface in the calculation of low energy electron diffraction experiment where measured electric current - voltage curve compared with the corresponding calculation. The smaller the value of R, the better the functions $f$ and $g$. Thus, we use a random number to search for the parameter set that  minimizes R. This random number technique is similar to the Metropolis method\cite{Metropolis}, which we have used previously\cite{Ishii2012a}. We use this R-factor as a guide to obtain the best parameters for each calculation in this paper.

We employ the Monte Carlo method like Metropolis method\cite{Metropolis} to fine the minimum of R. This is very similar for finding the minimum of total energy in the first principle calculation.  In the real calculation to adjust the parameters $C_x$, $D$, and $P$, we should take care of the local minimum trapping like the first principle calculation in material physics. It is well-known that there are several ways to find the minimum condition like the steepest descent, the equation of motion method and the conjugate gradient method. Even in the actual calculation of the first principle calculation or the density functional theory, we should be careful of the danger of local minimum trapping. In this paper, the way we employ is just do the calculation using the several initial value in the Metropolis-like method to avoid the local minimum trapping. To check the accuracy of the parameters adjusting, we use the R-factor value. For every calculation which we show in this paper, the R-factor is below 0.01. 

Actually, the parameters $C_{TV}$, $C_{NetNews}$, $C_{Twitter}$, $C_{blog}$, $D$ and $P$ in equation (\ref{eq:eq14}) can be considered as functions of time, because people's  attention changes over time. However, if  we introduce the functions $C_{TV}(t)$, $C_{NetNews}(t)$, $C_{Twitter}(t)$, $C_{blog}(t)$, $D(t)$ and $P(t)$, we can tune any phenomena by adjusting these functions. Thus, we retain $C_{TV}$, $C_{NetNews}$, $C_{Twitter}$, $C_{blog}$, $D$ and $P$ as constant values to examine whether equation (\ref{eq:eq14}) can be applied to any social phenomena.

\section{Data}

For the investigation of this article, we should use Google Trends data for the target data of our calculation.  The data of Google Trends can be obtained on the Google Trends page. 

The daily posting number to Twitter and Blog are obtained from the service "Kuchikomi Kakaricyo" of Hottolink co.ltd. The mass media advertisement data can be obtained from M Data co.ltd. via the "Kuchikomi Kakaricyo" service.

\section{Result}

Analyze the events 1 month before and 1 month after the events, and calculate each coefficients $C_{TV}$, $C_{NetNews}$, $C_{Twitter}$, $C_{blog}$, $D$ and $P$  so that the calculated $I(t)$ is combined with the measured value by Google Trend to best match.

\begin{figure}[h]
\begin{center}
\includegraphics[width=9cm]{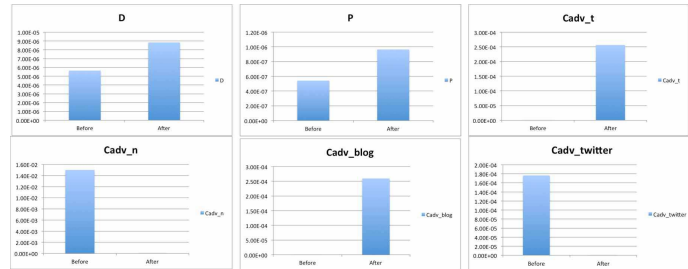}
\caption{Result of $D$, $P$, $C_{TV}$, $C_{NetNews}$, $C_{blog}$, $C_{twitter}$ in calulation for Christmas.}
\label{christmas}
\end{center}
\end{figure}

The calculation results of all coefficients for Christmas are shown in the fig.\ref{christmas}. For Christmas, the calculation results of all coefficients for the first month before Christmas and one month after Christmas are shown in the fig.\ref{christmas}.

\begin{figure}[h]
\begin{center}
\includegraphics[width=9cm]{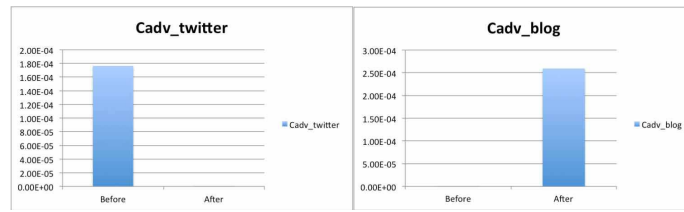}
\caption{Result of calulation of $C_{Twitter}$ and $C_{blog}$ before and after Christmas.}
\label{christmas-blogtwitter}
\end{center}
\end{figure}

In particular, here we examine the difference whether search behavior is affected by Blog or influenced by Twitter. Shown in the fig.\ref{christmas-blogtwitter} are $C_{blog}$ and $C_{twitter}$ values before and after Christmas. Looking at the results, Twitter's influence is big before Christmas and blog influence is big after Christmas.

\begin{figure}[h]
\begin{center}
\includegraphics[width=9cm]{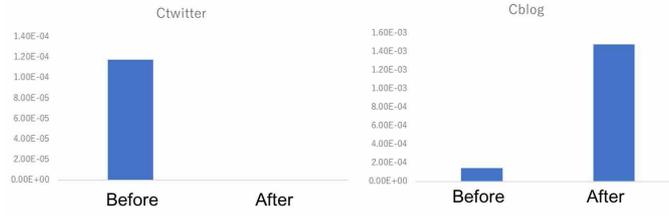}
\caption{Result of calulation of $C_{Twitter}$ and $C_{blog}$ before and after Halloween.}
\label{halloween-blogtwitter}
\end{center}
\end{figure}

\begin{figure}[h]
\begin{center}
\includegraphics[width=9cm]{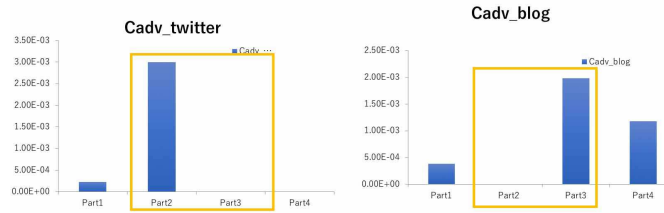}
\caption{Result of calulation of $C_{Twitter}$ and $C_{blog}$ before and after New Year Countdown. The analysis is done for a week before the Christmas, a week before the New Year's Day, a week after the New Year's Day and a week after January 8. }
\label{countdown-blogtwitter}
\end{center}
\end{figure}

\begin{figure}[h]
\begin{center}
\includegraphics[width=9cm]{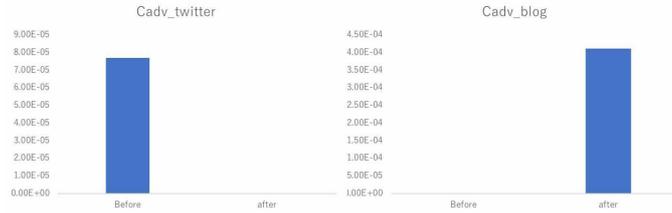}
\caption{Result of calulation of $C_{Twitter}$ and $C_{blog}$ before and after Valentine's Day.}
\label{valentine-blogtwitter}
\end{center}
\end{figure}

In the same way, the results calculated for Halloween, New Year's countdown, Valentine's Day are shown in the fig.\ref{halloween-blogtwitter}, fig.\ref{countdown-blogtwitter} and fig.\ref{valentine-blogtwitter}, respectively. For Holloween and Valentine's Day, we analyze the events 1 month before and 1 month after the events. For New Year's countdown, we analyze the event 1week before and 1 week after the event in order to avoid the effect of the Christmas before Dec.24. From these, it is possible to read the common tendency that Twitter influences strongly before the event, and after the event the effect of Blog is strong.

As mentioned above, qualitatively, Halloween, Christmas, Countdown and Valentine's Day are in agreement that it is influenced by Twitter before the event and after the event is affected by Blog.

\begin{figure}[h]
\begin{center}
\includegraphics[width=9cm]{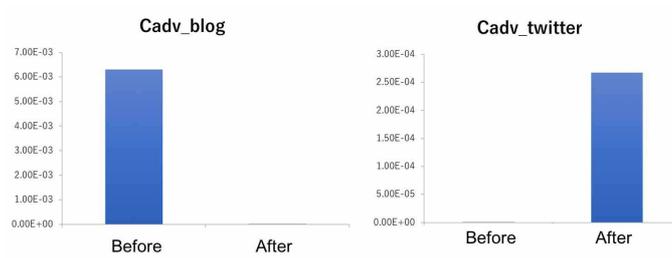}
\caption{Result of calulation of $C_{Twitter}$ and $C_{blog}$ before and after the Eho-maki.}
\label{eho-maki-blogtwitter}
\end{center}
\end{figure}

\begin{figure}[h]
\begin{center}
\includegraphics[width=9cm]{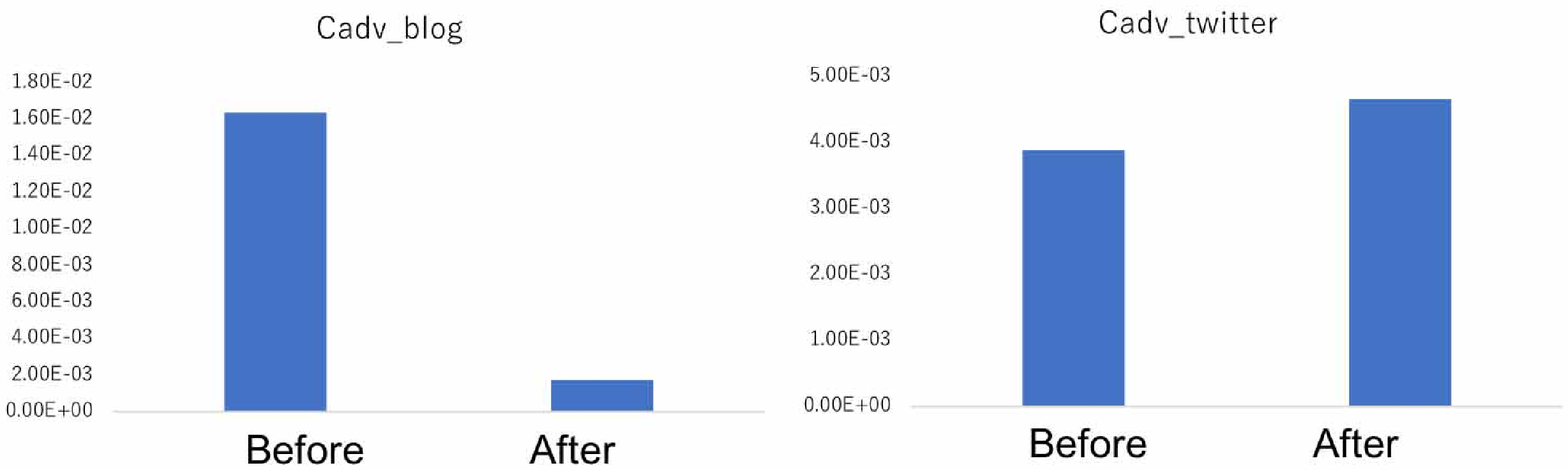}
\caption{Result of calulation of $C_{Twitter}$ and $C_{blog}$ before and after the day of the ox in midsummer.}
\label{unagi-blogtwitter}
\end{center}
\end{figure}

Next, we introduce the calculation results of Eho-maki and Eel of the Midsummer's Day eating foods decided on a fixed date in fig.\ref{eho-maki-blogtwitter} and \ref{unagi-blogtwitter}. As you can see from the results, the influence of Blog and Twitter on people's search behavior is opposite to the previous examples. In the case of this foods, it is from Blog that is influenced before the event and it is influenced by Twitter after the event. The qualitative behavior is same for the two event foods.

\section{Discussion}

We analyzed Christmas, Halloween, New Year Countdown, Valentine's Day, Ewaki Roll,  and Eel of the Midsummer Day of the ox as events with limited time. Looking at the analysis results, the results were divided for four of Christmas, Halloween, Countdown, Valentine's Day, and the two of Eho-maki and Eel. 

These two groups are thought to depend on surprises or happening for the event and its preparation. In Halloween, there are many matters to investigate beforehand, such as what kind of costumes they themselves, what kind of costumes they are going to do, what kinds of costumes will be popular this year. At Christmas in Japan, there are not many things to consider beforehand, such as what kind of surprise there are lovers to book a Christmas dinner with. In the New Year 's countdown, since the taste of the countdown is different for each gathering, it is necessary to gather information in advance according to which place to go to, which event to go to and which bar to go to.  At Japanese Valentine's Day, women collect information in advance, whether women make their own chocolate for her lover or they purchase high-end chocolate at some famous brand shop. 

Meanwhile, there are few kinds of Eho-maki to eat as an event, and there is no element to look into in advance as there are also decided how to eat. Also, there is no surprise when eating Eho-maki. As for the Eel to eat on the Midsummer Day of the ox, as shown in the fig.\ref{unagi}, the method of cooking has been decided traditionally. It is impossible to make home made, so Japanese people have to eat Eel at a restaurant. Therefore, the information to be checked in advance is the only restaurant to eat. Besides, there is not much difference in Eel's cuisine for each restaurant.

In this way, events that need to check information sufficiently in advance are affected by Twitter. On the contrary, in the case of an event where there is no surprise and there is no need to check in advance, the effect of Twitter at the prior stage is small, and the influence of Blog is expected to be relatively high. 

Therefore, if search actions that are strongly influenced by Twitter are observed beforehand, those who come to the event are collecting information. For those people, event-related marketing will be effective. In this way, the method of this research can be applied to marketing.

\section{Conclusion}

Posting on Blog and Twitter is an act performed by some people in society. On the other hand, search behavior is used by most people who use the Internet. Therefore, analysis of search behavior on the Internet is very important in social analysis in that it can also target people without voice.

In this research, we have analyzed such search behavior on the Internet using mathematical model of search behavior. As an object, we observed preliminary excitement and post cool down at the event to be held in a short time. The events analyzed are Halloween, Christmas, Countdown, Valentine's Day, Eho-maki, Eel of the Midsummer Day of the ox in Japan. According to the analysis, it turned out that these are divided into events requiring preparatory preparations for surprises, and events not being prepared. Twitter has a strong influence on events requiring advance preparations for surprises. In other events the effect of Twitter is after the incident.

This research result is expected to be applicable to marketing.

\section*{Acknowlegement}

The authors are grateful for helpful discussion to Yasuko Kawahata of Gunma University, Japan.

\end{document}